\documentclass{amsart}

\usepackage[utf8]{inputenc}
\usepackage{amsmath}
\usepackage{mathtools}
\usepackage{amsaddr}
\usepackage{graphicx}
\usepackage{amssymb}
\usepackage{mathrsfs}
\usepackage[nobysame, alphabetic]{amsrefs}
\usepackage{hyperref}
\usepackage[margin=1.5in]{geometry}
\usepackage{xcolor}

\newtheorem{definition}{Definition}

\newtheorem{remark}{Remark}

\newtheorem{problem}{Problem}
\newtheorem*{conjecture*}{Conjecture}
\newtheorem*{question*}{Question}
\newtheorem*{theorem*}{Theorem}

\numberwithin{definition}{section}

\newcommand{\NN}{\mathbb{N}}

\newcommand{\cA}{\mathcal{A}}

\renewcommand{\epsilon}{\varepsilon}

\def\(#1\){\begin{align*}#1\end{align*}}

\title{On the use of Dynamical Systems in Cryptography}
\author{Samuel Everett}
\email{same@uchicago.edu}
\keywords{Chaotic dynamical system, Discrete dynamical systems, Cryptography, Encryption, Stream cipher, One-way functions}
\subjclass[2010]{37N99, 37A50, 94A60, 94A62}

\begin{document}

\begin{abstract}
Ever since the link between nonlinear science and cryptography became apparent, the problem of applying chaotic dynamics to the construction of cryptographic systems has gained a broad audience and has been the subject of thousands of papers.  
Yet, the field has not found its place in mainstream cryptography, largely due to persistent weaknesses in the presented systems.  The goal of this paper is to help remedy this problem in two ways.  The first is by providing a new algorithm that can be used to attack -- and hence test the security of -- stream ciphers based on the iteration of a chaotic map of the interval.  The second is to cast discrete dynamical systems problems in a modern cryptographic and complexity theoretic language, so that researchers working in chaos-based cryptography can begin designing cryptographic protocols that have a better chance of meeting the extreme standards of modern cryptography.
\end{abstract}

\maketitle

\section{Introduction}
The field of chaos-based cryptography is concerned with the design of private-key cryptographic protocols based on chaotic dynamical systems.  Due to the statistical properties of chaotic dynamical systems, as well as the rich body of literature distinguishing the complex nature of such systems \cite{devaney2021introduction,katok}, it is natural to attempt to use these systems in the development of various cryptographic protocols.

Indeed, since the pioneering work of Matthews, Baptista, and Kultulski et al. \cite{matthews1989derivation,Baptista,kotulski1} describing how a cryptosystem can be constructed using chaotic maps of the interval, there have been thousands of publications following suit, exploring the myriad ways in which one can construct cryptographic primitives and protocols using chaotic dynamical systems.  However, despite the incredible attention chaos-based cryptography has received over the years, a well known problem the field faces is its inability to capture the attention of cryptographers and computer-scientists actually building and implementing cryptosystems (see \cite{amigo2009chaos,solak,solak2011} for discussion).  In fact, since the publication of Baptista's original paper twenty-four years ago, only one paper describing the use of chaotic dynamics in cryptography has appeared in any of the primary cryptography conferences \cite{habutsu1991secret}, and in fact that cryptosystem was broken the very same conference \cite{biham1991cryptanalysis}.

Although researchers working in chaos-based cryptography have established a number of strong results (see e.g. \cite{carmen,iyengar2018q,kocarev,schmitz2001,dachselt1998} for review), the lack of attention from the main-stream cryptographic community is not unjustified: the frequent publication of insecure cryptosystems is endemic in the chaos-based cryptography literature, leading to an exceptionally scarce selection of cryptographic protocols that appear to actually be secure.
Indeed, it is common to see a proposed chaos-based cryptographic scheme get broken shortly after publication (see e.g. \cite{li2002cryptanalysis,liu2020cryptanalysis,hu2017cryptanalysis,wu2023cryptanalysis,zhang2013cryptanalysis,he2012cryptanalysis}), and -- as this paper will demonstrate -- many protocols that have not been successfully cryptanalyzed do not necessarily remain as such because they are secure, but because apparently there has been little research done in attempting to break these systems.

Moreover, there have been a number of widely read papers in the discipline that attempt to provide frameworks and rulesets designers of chaos-based encryption schemes can use as a reference to assess the security of their protocols (\cite{dachselt,amigo,amigo2009chaos,carmen,alvarezlessons,schmitz2001}).  However, while these papers provide an important step in the right direction, they tend to emphasize evaluating the security of protocols by studying the statistical properties of the systems, ensuring that they pass batteries of statistical tests, and that the underlying dynamical systems (before being discretized) have characteristics like topological mixing and positive Lyapunov exponents.  While these guidelines describe features any cryptosystem must exhibit, they are not sufficient, and generally fail to align with modern notions of security established in the cryptography community, which focus on rigorously proving the security of protocols conditioned on the assumption that some well-understood problem has no efficient algorithm solving it.

To remedy this issue, it is necessary that the chaos-based cryptography community changes its standards for evaluating the security of protocols to match the stringent and rigorously defined methods used in modern cryptography.
There have been a number of important works to this end \cite{teh2020implementation,arroyo2017cryptanalysis,alvarezlessons,dachselt,amigo}, as well as papers following the view of modern cryptography that study must be focused on designing one-way functions from which cryptosystems may be developed \cite{tutueva2020construction}.  
Nonetheless, there remains substantial work that must be done before the field of chaos-based cryptography can meet the quality standards of modern cryptography.
The aim of this paper is to help the field of chaos-based cryptography close this gap.

This paper accomplishes this aim in a number of ways.  First, in Section \ref{sec1}, we provide the basic background grounding modern cryptography, which researchers working in chaos-based cryptography ought to be aware of and work with.  In particular, we focus our attention on the definitions of pseudo-random generators, one-way functions, and basic security definitions and construction methodology for encryption protocols.  We make the case that the most productive direction the field of chaos-based cryptography can work toward is finding computationally intractable problems in discrete dynamical systems, which may not only provide an alternative to well-known basic intractable problems in number theory and algebra, but that could also illuminate deeper mathematical structures in the dynamical systems themselves.

Toward this goal, in Section \ref{sec2}, we introduce an algorithm designed specifically for cryptanalyzing stream ciphers built off of the iteration of chaotic maps of the interval.  We demonstrate how the algorithm we introduce can efficiently break a number of well-known stream ciphers in the chaos-based cryptography literature, and hence provides a powerful tool for assessing the security of chaos-based cryptosystems.  Finally, in Section \ref{sec3} we introduce a possibly computational intractable problem from discrete dynamical systems, which appears to be secure against known attacks, including that introduced in Section \ref{sec2}, and is hence worthy of further study.

\subsection*{Acknowledgments}
The author would like to thank David Cash for the support and feedback.  This material is based upon work supported by the National Science Foundation Graduate Research Fellowship Program under Grant No. 2140001. Any opinions, findings, and conclusions or recommendations expressed in this material are those of the author(s) and do not necessarily reflect the views of the National Science Foundation.

\section{Background and basic cryptographic definitions}\label{sec1}

Although pseudorandom number generators constructed using chaotic dynamical systems often provide desirable statistical properties \cite{kohda1997statistics}, contrary to the ideas propagated in a number of widely read articles (see e.g. \cite{alvarez,amigo,ozkaynak,kocarev,schmitz2001}), by modern standards a cryptographic protocol cannot be deemed secure if it passes a battery of statistical tests and is resistant to well-known attacks.  Such tests are \textit{necessary} but by no means \textit{sufficient} in determining the security of cryptosystems.  Rather, modern cryptography places heavy emphasis on definitions, precise assumptions, and rigorous proofs of security. 

More specifically, modern cryptography is built off the recognition that formal definitions of security are essential in the design of cryptographic primitives and protocols.  In fact, cryptographic definitions of security are quite strong and not easily achievable. The protocols that do meet such rigorous security definitions often cannot be proven secure in an unconditional sense, but instead rely on some widely-believed (but unproven) assumption, such as the assumption that no efficient algorithm exists for finding prime factors of large integers.  And, importantly, it is taken that such assumptions are clearly defined, thereby enabling rigorous security reductions.

This point provides the essence of modern cryptography: \textit{cryptographic constructions can be proven secure with respect to a security definition and relative to a well-defined assumption}.  This is the key feature which separates almost all of modern chaos-based cryptography research from mainstream cryptography research; namely, chaos based cryptography research still stands firmly in the realm of ``classical cryptography," where cryptographic schemes are deemed secure if the designers of said scheme could not break them.  In contrast, modern cryptography employs mathematical proofs of security so that cryptographic schemes are guaranteed to be secure unless the underlying assumption is false.  Moreover, the underlying assumptions are often long-standing and well studied.  This gives the designers and users of the protocol confidence that the scheme is secure, because great effort has been unsuccessful in efficiently solving the underlying problem.

In the following, we provide key definitions and notions critical in the study of cryptography, of which any researcher developing dynamical systems-based cryptosystems should be familiar and refer to.  The reader is encouraged to read parts I \& II of \cite{katzlindell} or Chapters 1--3 of \cite{goldreich} for a broader and more detailed treatment of the following material.  We also refer the reader to \cite{boneh2020graduate} for a more applied perspective.

We begin by summarizing three basic principles of modern cryptography \cite{katzlindell}:
\begin{enumerate}
    \item \textit{Principle 1:} The first step in solving any cryptographic problem is the formulation of a rigorous and precise definition of security.
    \item \textit{Principle 2:} When the security of a cryptographic construction relies on an unproven assumption, this assumption must be precisely stated.  Furthermore, the assumption should be as ``minimal" as possible, and should be well-known and widely studied.
    \item \textit{Principle 3:} Cryptographic constructions should be accompanied with a rigorous proof of security with respect to the definition formulated according to principle 1, and relative to an assumption stated as in principle 2.
\end{enumerate}
In addition to the central principles of modern cryptography, another important aspect is the move toward security definitions based on the idea of \textit{computational security}.  That is, modern encryptions schemes have the property that they can be broken given enough time and computation, but under the right assumptions, the amount of computation required to break these schemes is intractable.  In particular, the computational approach incorporates two assumptions:
\begin{enumerate}
\item Security is only preserved against \textit{efficient adversaries} (adversaries with limited time and compute power), and,
\item adversaries can succeed in breaking the encryption scheme with \textit{negligible probability}, which is equated with success probabilities smaller than any inverse polynomial in $n$, where $n$ is taken to be a \textit{security parameter} chosen when generating keys (usually taken to be the number of bits in the secret key).
\end{enumerate}
The second point highlights the emphasis of modern cryptography around an ``asymptotic approach," where security of a system is guaranteed only for sufficiently large security parameters (which is usually key size).

We view the running time of the adversary and the honest parties as a function of the security parameter $n$, and we assume the adversary knows the value $n$. We now give the definition of polynomial time algorithms, providing the notion of ``efficient computation" which is foundational in modern cryptography and complexity theory.

\begin{definition}
An algorithm $A$ is said to run in \emph{polynomial time} if there exists a polynomial $p(\cdot)$ such that for every input encoded as a binary string $x \in \{0, 1\}^*$, the computation of $A(x)$ terminates within at most $p(||x||)$ steps, where $||x||$ denotes the length of string $x$.  

A \emph{probabilistic algorithm} is an algorithm that has the capability of tossing coins: it has access to a random source that returns uniformly distributed random bits.  A probabilistic polynomial time algorithm (PPT) algorithm $A$ is a polynomial time algorithm with access to a random source.
\end{definition}

We also formalize the notion of a negligible function.

\begin{definition}
A function $f$ is \emph{negligible} if for every polynomial $p(\cdot)$ there exists an $N$ such that for all integers $n > N$ it holds that $f(n) < 1/p(n)$.
\end{definition}

We are now prepared to give our primary definitions.  Notably, we choose not to give security definitions for symmetric-key (private-key) cryptosystems (see Part I of \cite{katzlindell} for such definitions).  Rather, we deliberately choose to focus solely on the definitions of pseudorandom generators (PRGs) and one-way functions (OWFs).  
Our primary reason for this choice is that the design of such cryptographic primitives is hard and theoretically interesting, and finding fundamentally hard problems in dynamical systems could not only have application to cryptography, but could very well reveal deep features about the dynamical systems themselves.  In addition, the reason why most chaos-based cryptosystems are so often and so easily broken is because the underlying problem is not actually difficult to solve.  So, a highly productive direction would be for the community to spend its time finding intractable problems in dynamical systems from which cryptosystems could be developed.

\subsection{Pseudorandom generators.}  
A pseudorandom generator is a deterministic algorithm that receives a short truly random seed and stretches it into a long string that ``looks random" to efficient distinguishers.

\begin{definition}[Pseudorandom generator (PRG)]\label{defPseudoGenerator}
Let $l$ be a polynomial and $G$ be a deterministic polynomial-time algorithm such that upon any input $s \in \{0, 1\}^n$, algorithm $G$ outputs a string of length $l(n)$.  We say that $G$ is a \textsf{pseudorandom generator} if the following two conditions hold:
\begin{enumerate}
    \item Expansion: for every $n$ it holds that $l(n) > n$.
    \item Pseudorandomness: For all probabilistic polynomial-time distinguishers $D$, there exists a negligible function $\emph{\textsf{negl}}$ such that
    \begin{equation}
    |\emph{Pr}[D(r) = 1] - \emph{Pr}[D(G(s)) = 1]| \leq \emph{\textsf{negl}}(n)
    \end{equation}
    where $r$ is chosen uniformly at random from $\{0, 1\}^{l(n)}$, the seed $s$ is chosen uniformly at random from $\{0, 1\}^n$ and the probabilities are taken over the random coins used by $D$ and the choice of $r$ and $s$.
\end{enumerate}
The function $l$ is called the \textsf{expansion factor} of $G$.
\end{definition}
Intuitively, this definition states that $G$ is a PRG if it expands its input, and if there exists no efficient algorithm that can distinguish the output of $G$ from the output of a truly random source.  An alternative characterization of a PRG, attributed to Yao \cite{yao1982theory}, states that a generator is a PRG if any PPT attacker who knows the first $i$ bits of output (but not the seed, and $i$ is polynomial in $n$) has negligible advantage in predicting the $(i+1)$st bit.

Although the existence of PRGs is not known, under the assumption they exist we can construct provably secure cryptosystems, by \textit{reducing} the security of the cryptosystem to the assumption that there exists a PRG.  For instance, the natural construction where the output of a PRG is XORed with a plaintext message can be proved secure under the assumption that there exists a PRG, and with respect to specific definitions of security (see Chapter 3 of \cite{katzlindell} for details).

\subsection{One-way functions.}
As we have discussed up to this point, one fundamental tenet of modern cryptography contrasting it with classical cryptography is that cryptographic protocols are \emph{proven} to be secure, typically conditioned on an assumption.  In addition, a natural goal is to make this assumption as ``minimal" as possible.  That is, that the assumption is simple, broadly studied, and widely believed to be true. Formally, a minimal assumption is one that is both necessary and sufficient for achieving constructions of pseudorandom generators, and pseudorandom functions and permutations (block ciphers).  Although the definitions given here emerged from years of research, see the seminal work \cite{goldreich1986construct} and the references therein for the introduction of and discussion of basic cryptographic notions.

The most minimal assumption applied in modern cryptography is that of a \emph{one-way function}, which is a function having the property that it is easy to compute, but hard to invert (almost always).  That is, a function is one-way if it can be computed by a polynomial time algorithm, but no probabilistic polynomial time algorithm can compute its inverse (find a preimage), except with negligible probability.

\begin{definition}[One-way functions (OWF)]\label{defOWF}
A function $f:\{0, 1\}^* \rightarrow \{0, 1\}^*$ is called \textsf{one-way} if the following two conditions hold:
\begin{enumerate}
\item \emph{Easy to compute:} There exists a polynomial-time algorithm $M_f$ such that on any input $x \in \{0, 1\}^*$, $M_f$ outputs $f(x)$.
\item \emph{Hard to invert:} For every probabilistic polynomial time inverting algorithm $\cA$, there exists a negligible function $\emph{\textsf{negl}}$ such that
\begin{equation}
Pr[\cA(f(x)) \in f^{-1}(f(x))] \leq \emph{\textsf{negl}}(n)
\end{equation}
where the probability is taken over the uniform choice of $x$ in $\{0, 1\}^n$ and the random coin tosses of $\cA$.
\end{enumerate}
\end{definition}
We stress that it is only guaranteed that a one-way function is hard to invert when the input is \textit{uniformly distributed}.  Thus, there may be many inputs for which the function can be inverted (but this is a negligible fraction).
Moreover, a one-way function can be inverted given enough time: given a value $y$, it is always possible to try all values of $x$ of increasing length until a value of $x$ is found such that $f(x) = y$.  Thus, the existence of OWFs is inherently an assumption about computational complexity and hardness.  

In particular, we note that if there exists one-way permutations (bijective OWFs), then there exist pseudorandom generators.  And, equipped with a one-way function, we can develop a wide variety of provably secure encryption schemes.  For this reason, one-way functions are quite fundamental, and given the complicated nature of discrete chaotic dynamical systems, it appears that attempting to find computational intractable problems in dynamical systems which can be converted into one-way functions may be fruitful, and is likely the best direction the field can orient itself to.

Indeed, rather than assuming some chaos-based encryption scheme is secure -- which is highly dangerous and summarizes the current state of chaos based cryptography -- it would be better to instead assume that some ``small," widely studied low-level problem coming from dynamical systems is hard to solve, and then prove that the construction in question is secure given this assumption.  This is a desirable approach because then the same basic assumptions can be widely applied to a variety of schemes, and the assumptions are more likely to be true because smaller problems are easier to understand and find flaws in.

The proof that a given construction is secure as long as some underlying problem is hard generally proceeds by presenting an explicit reduction showing how to convert any efficient adversary $\cA$ that succeeds in breaking the construction with non-negligible probability, into an efficient algorithm $\cA'$ that succeeds in solving the problem that was assumed to be hard.  That is, assuming some problem $\textsf{X}$ cannot be solved by any polynomial time algorithm except with negligible probability, we want to prove that some cryptographic construction $\Pi$, like a stream or block cipher, is secure in the sense that an algorithm breaking the scheme can be used to solve the underlying problem $\textsf{X}$, using $\cA$ breaking the scheme as a black-box.  It is recommended that the reader read the first part of \cite{katzlindell} to learn the details concerning proofs of security and reductions.

\subsection{The importance of modern security definitions.}
The key takeaway from this section is that modern cryptosystems must be proven secure, with a reduction to a well-defined, widely studied problem that is considered to be hard to break.  Moreover, it is important to note that modern security definitions in use today are extremely rigorous and have been refined over decades to give what we believe are good security guarantees.

As a consequence, the pervasive statistical analysis techniques used in the chaos-based cryptography literature to determine the security of cryptosystems are in fact wholly inadequate by modern standards.  Indeed, if a cryptosystem reduces to a problem not known to be solvable in polynomial time, then it will pass all statistical tests, but passing statistical tests does not imply the problem is hard to solve, as will be demonstrated in Section \ref{sec2}.

To this end, future work on chaos-based cryptosystems should focus on reducing their security to well-known problems in dynamical systems that are deemed to be computationally intractable (see Section \ref{sec3} for a candidate problem).

\section{An attack on chaos-based stream ciphers}\label{sec2}

We now move to outline an algorithm attacking a class of stream ciphers coming from the iteration of a chaotic map of the interval.  The basic idea of the algorithm is straightforward: given a stream of encrypted output from the cipher, along with a block of plaintext, we obtain the stream of bits from the PRG underlying the stream cipher (the output used to encrypt the plaintext).  We then use this output to iteratively shrink a candidate set of initial conditions (keys/seeds of the stream cipher keying the PRG) down until we have a unique result, or a small set of candidates that can be efficiently brute-forced.  The algorithm presented here is similar in spirit to the bisection method used for root finding -- a useful analogy to keep in mind.

More precisely, the fundamental logic the algorithm is based on is as follows.  Take a segment of output $a_ia_{i+1}\cdots a_{i+t}$ from the PRG underlying the stream cipher, which can be obtained using a segment of plaintext -- something any reasonable security model will allow the adversary to have access to.  The basic step to take is the following: given a target output value $a_i$ of the PRG at it's $i>0$th iteration,  determine the set of seeds $S_i$ that satisfy the condition that $G^i(s) = a_i$, where $s\in S_i$ is a seed and $G$ a PRG.  If this set $S_i$ of seeds is efficiently computable, then upon determining the set $S_i$, we can ask: what is the subset $S_{i+1} \subset S_i$ of seeds satisfying the condition $G^{i+1}(s) = a_{i+1}$ for $s \in S_{i+1}$?  Continuing in this manner, we obtain a sequence of subsets $S_{i+t} \subset S_{i+t-1} \subset \cdots S_i$, with $S_{i+t}$ very small in cardinality if $t$ is sufficiently large, thereby allowing the set of candidate secret seeds to be brute-force checked.

This logic is clearly very general, and is the reason why the algorithm presented below can be applied to a wide variety of chaos-based cryptosystems; in practice, the algorithm only uses very basic mathematical structure pervasive in chaos-based cryptosystems.  To this end, although the algorithm given below is stated only in terms of stream ciphers built off PRGs coming from iterating a single chaotic map of the interval, it is nevertheless readily generalizable to a much broader class of chaos-based cryptosystems.  Indeed, exploring such generalizations and developing the associated rigorous algorithms would be a fruitful and impactful line of study.

The kinds of schemes this algorithm attacks were thought to be secure because the underlying problems were considered complex (see discussion in \cite{schmitz2001}).  However, the attack described here contradicts this notion, and in fact can be used to break a number of chaos-based cryptosystems (see e.g. \cite{al2012new,li2017image,li2006multiple,wang2016pseudorandom,xiang2006novel}).  In fact, we remark that it has been known for some time that cryptosystems based on iterating a single chaotic map are insecure because there are a number of tools from dynamics that can extract information from such schemes \cite{li2001pseudo}.  Strengthening these findings, the attack given in this paper completely breaks many such schemes, and hence this result taken in conjunction with previous results is a strong indicator that cryptosystems based off the iteration of a single chaotic map should not be considered for use in cryptography.

\subsection{An efficient algorithm breaking PRGs based on iteration of a chaotic map of the interval.}

Let $n$ denote the security parameter, $\Sigma$ denote a finite alphabet with cardinality at least two, and $I=[0,1]$ denote the unit interval.  Let $f:I\rightarrow I$ be a unimodal, continuous chaotic map of the interval, and $g:I\rightarrow \Sigma$ a step function carrying points from $I$ to the finite alphabet $\Sigma$.  We let $G(k)$ label a pseudorandom generator (PRG), defined so that given $k \in I$, $G^i(k)\in \Sigma$ is the $i$th symbol output for $i \in \NN^+$.  Define the iterates of $G$ so that $G^i(k) \vcentcolon= g(f^i(k))$.  $G(k)$ may be used as a stream cipher, so that for secret key $k \in I$, the iterates $G^i(k) \in \Sigma$ constitute the output of the PRG which is used to encrypt a plaintext $p$ in some way.

Additionally, we remark that $f$ and $g$ must be computable in time polynomial in $n$.  As an example, $g$ cannot be a step function with an infinite description, or one with a description somehow an exponential function in $n$.  Generally, the requirement that $f$ and $g$ must be computable in polynomial time (i.e. in the class \textsf{FP}) restricts their complexity.
In particular, we note that following the restrictions on $f$ and $g$, if $A \subset I$ is the set of points such that $G(x) = \sigma \in \Sigma$ for all $x \in A$ and fixed $\sigma \in \Sigma$, then it follows that $A = \sqcup_{i=1}^s A_i$ where the $A_i$ are intervals, and $s = poly(n)$.

Under this model, the task of an adversary is to determine the secret initial condition $k \in I$ given some contiguous subsequence of output from the PRG underlying the stream cipher:
\begin{equation}
C = \{G^i(k)\}_{i=m}^{m+t} \subseteq \{G^j(k)\}_{j=1}^\infty,
\end{equation}
where $m \geq 1$, $t = poly(n) > n$.  \emph{As a consequence, for the adversary to obtain such output it is necessary they have access to some contiguous block of plaintext of length $t$}.  Moreover, by Kerckhoffs's principle we assume the adversary has complete knowledge of $f, g$, $I$ and $n$.  Let $C = \{G^i(k)\}_{i=m}^{m+t} = \{c_i\}_{i=1}^{t}$ so that $c_i$ labels the $i$th symbol returned from the PRG that the adversary has access to.

\begin{remark}
In the following we are describing an algorithm implemented on a computer, and as a consequence when we speak of ``intervals" and ``functions on intervals," we are implicitly speaking of the discretized version of the underlying dynamical system, where intervals are thought of as contiguous sequences of points contained in the dyadic rationals $I = \{i/2^n\}_{i=0}^{2^n}$ where $n$ is the security parameter, because the secret will be a random point in $I$.

As a consequence, for the purposes of this paper we operate under the fixed point arithmetical setting with rounding.  In particular, the algorithm given here supposes use of the techniques of interval analysis and outwards rounding to obtain accurate results in working with intervals \cite{moore2009introduction,a7140721}.  In doing so we concretely tie the security parameter $n$ to the length of the secret key, making this formalism rigorous and aligned with the theoretical computational complexity setting.  Further development and study of the kind of algorithm given here under floating point arithmetic would constitute an important line of study (see e.g. \cite{overton2001numerical}).
\end{remark}

Consider the following algorithm breaking the chaos-based pseudorandom generator outlined above.
\begin{quote}
\textbf{Algorithm.}  On input of security parameter $n$, $\Sigma$, the set $C=\{c_i\}_{i=1}^{t}$ of output from $G$ obtained by using a length $t$ plaintext block, with $t = poly(n)$, and a description of $f$ and $g$, proceed as follows.
\begin{enumerate}
\item Given $c_1 \in \Sigma$, let $A^{(1)} = \sqcup_{i=1}^s A_i^{(1)}$ label the disjoint union of intervals $A_i^{(1)}\subset I$ such that $G[A_i^{(1)}] = c_1$.  Each interval $A_i^{(1)}$ composing $A^{(1)}$ can be encoded as a pair $(a, b)$ of numbers.  Compute the set $E_1$ of $s$ endpoint pairs determining the intervals $A_i^{(1)}$.

The collection $E_1$ can be efficiently computed by using a root finding algorithm to find all pairs of points $a, b \in I$ such that $g(f[(a,b)]) =c_1$ as follows.  $g$ is taken to be a step function mapping into $\Sigma$.  As such, using the description of $g$, read off the endpoint values determining intervals $I' \subset I$ for which $g[I'] = c_1$.  Suppose $w_1,...,w_q$ with $q$ even are these interval endpoint values, so that $(w_\alpha, w_{\alpha+1})$ compose intervals such that $g[(w_\alpha, w_{\alpha+1})] = c_1$, with $\alpha = 1,3,...,q-1$ taken to be odd.  Then, for each $w_\alpha$, compute the roots $r_i$ of $f(x) - w_\alpha$.  The total number of such roots must be even since $f$ is unimodal, so the found roots can then be joined into pairs corresponding to each $A^{(1)}_i$ interval, by ordering the roots and pairing off consecutive values (i.e. if $(r_1<r_2 <\cdots <r_p)$ then $(r_1,r_2),(r_3,r_4),...,(r_{p-1},r_p)$ are the $A_i^{(1)}$).  These pairs $(r_i, r_{i+1})$ then compose the elements of $E_1$, since for each pair $(r_i,r_{i+1})$ we have $g(f[(r_i,r_{i+1})]) = c_1$\footnote{For example, if $g$ is the step function rounding to the nearest integer, and $c_1=0$, compute the roots of $f(x)-0$ and $f(x)-1/2$.  Then, these roots constitute the endpoints of the $A_i^{(1)}$ by ordering the roots and pairing off the four roots into two pairs.  Steps of this process are demonstrated in Section 3.2.}.

\item In the $i$th step: given symbol $c_i$, $i\leq t$, and a set of endpoint pairs $E_{i-1}$, for each pair $(a_j, b_j) \in E_{i-1}$, determine endpoints $a_j', b_j'$ such that $(a_j', b_j') \subsetneq (a_j, b_j)$ and $G^i[(a_j', b_j')] = c_i$ as above.  Note that there must always exist at least one such interval $(a_j', b_j')$ such that $(a_j', b_j') \subsetneq (a_j, b_j)$, due to the assumption that $f$ is chaotic (see Remark \ref{remark3}).

Repeat for every endpoint pair in $E_{i-1}$, and use the new endpoint pairs $(a_j', b_j')$ to construct a set of endpoint pairs $E_{i}$.  Repeat this step $t$ times.  Each $E_i$ can be computed using the same root-finding technique outlined in Step 1, where the root search for $f^i(x)-w_\alpha$ is restricted to the set $A^{(i-1)}$, with $A^{(0)} = I$.

\item By assumption, $t = poly(n) > n$, and hence it follows that for every interval pair $(a_j, b_j)$ in $E_t$, we have
\begin{equation}
\left| \left(\bigcup_j(a_j, b_j)\right) \cap \{i/2^n\}_{i=0}^{2^n}\right| = poly(n).
\end{equation}
That is, the number of candidate secret initial conditions $x \in I$ such that $\{G^i(x)\}_{i=1}^t = \{c_i\}_{i=1}^t$ is polynomial in the size of $n$ and can hence be efficiently brute force-checked.  It follows that upon finding the correct initial condition $x \in I$, the adversary knows that $\{G^i(x)\}_{i=1}^\infty = \{G^i(k)\}_{i=m}^\infty$, and hence knows all future output from the PRG, thereby breaking the scheme.  This constitutes a complete break of the PRG, and hence the corresponding stream cipher by modern cryptographic standards.

\end{enumerate}
\end{quote}

\noindent Some remarks are in order.
\medskip

\begin{remark}
An important note in regards to the efficiency of this algorithm is that the cardinality of the interval endpoint sets $E_i$ never exceeds $2s$ by the unimodality assumption of $f$, and hence the number of intervals that need to be dealt with at each iteration of the algorithm has a polynomial upper bound.
\end{remark}

\begin{remark}\label{remark3}
The claim in step 2 that $(a_j', b_j') \subsetneq (a_j, b_j)$ follows trivially from the fact that $f$ is taken to be chaotic (using Devaney's standard definition of chaos \cite{devaney2021introduction}).  In fact, it is clear that in this setting if $(a_j', b_j') \subsetneq (a_j, b_j)$ then $(b_j'-a_j')\leq C(b_j-a_j)$ for some fixed constant $C <1$, so the intervals containing candidate secret initial conditions shrink at an exponential rate as this algorithm progresses, which implies the correctness of step 3.
\end{remark}

As noted in the beginning of Section \ref{sec2}, the basic techniques this algorithm uses to attack PRGs based on iterating a chaotic map of the interval can indeed be extended to a broader class of chaos-based stream ciphers, such as those based on the iteration of higher dimensional dynamical systems like the H\'enon map (see 
\cite{benedicks1991dynamics,ibrahim2020efficient,suneel2009cryptographic,henon2004two} for review of the dynamics and examples of relevant cryptosystems), or the cases when the parameter of the chaotic map $f$ is kept secret as well.  However, such generalizations appear to require sophistication to be built correctly, and hence the generalization of the basic techniques given here constitutes an important and worthy line of study.

\subsection{An example application of the attack.}

In the following example we present a complete break of the ``tent map image encryption scheme" presented in \cite{li2017image}.  Although, we remark that the algorithm given above can be used to break a number of different schemes (see e.g. \cite{al2012new,li2017image,li2006multiple,wang2016pseudorandom,xiang2006novel}), the scheme given in \cite{li2017image} provides a particularly nice example due to the simple structure of the stream cipher.

\begin{remark}
Although the authors of \cite{li2017image} state their scheme is to be used for image encryption, modern cryptography makes no distinction between the types of data that are being encrypted.  This is because given a suitable encoder, any object can be encoded into a binary string, so modern cryptography concerns itself only with the encryption of binary strings.  
It is common for researchers in chaos-based cryptography to focus on the development of ``image encryption schemes," but this is ultimately meaningless, and rather researchers should focus on developing schemes that encrypt arbitrary binary strings.
\end{remark}

The scheme presented in \cite{li2017image} is fundamentally a stream cipher based on iteration of the tent map.  The tent map $f:[0, 1]\rightarrow [0, 1]$ is defined by
\begin{equation}
f(x) = \mu \min\{x, 1-x\}
\end{equation}
where $\mu \in [0, 2]$ is the parameter.  Depending on the value of $\mu$, $f$ may have orbits that are asymptotically stable, or chaotic.  For a detailed study of the behavior of the tent map see \cite{yoshida1983analytic}, or see \cite{holmgren2000first} for a broad overview and detailed analysis of ``equivalent" (topologically conjugate) systems.

The stream cipher described in \cite{li2017image} functions by fixing the parameter $\mu$ to be a value such that the system exhibits chaotic behavior, and takes the secret key to be an initial condition $x_0 \in [0, 1]$.  Simplifying, encryption of a plaintext (an image in the case of the paper) is then obtained by encoding the iterates $f^k(x_0)$ in binary, as well as the plaintext, and then XORing the two to obtain the ciphertext.  Although the paper leaves to interpretation exactly how the stream obtained by iterating the tent map is to be encoded into binary, for the purpose of this example take the encoding function $g:[0, 1]\rightarrow \{0, 1\}$ to be 
\begin{equation}
g(x)=
\begin{cases}
0 \text{ if } 0\leq x\leq \frac{1}{2},\\
1 \text{ if } \frac{1}{2} < x \leq 1.
\end{cases}
\end{equation}
As such, recasting the stream cipher in the notation used in the description of our attack, the stream cipher $G$ is defined as $G^i(x)=g(f^i(x))$, $i\in \NN$, so that with secret initial condition $x_0$, $\{G^i(x_0)\}_{i=1}^\infty$ is the output stream which is XORed with a plaintext message during encryption.

Suppose, for the sake of the following example attack, that the adversary has obtained an $m$-bit plaintext block and hence obtains some subsequence of the output stream $C = \{G^i(x_0)\}_{i=j}^{j+m} = \{c_i\}_{i=1}^m$ with $c_i \in \{0, 1\}$, and for security parameter $n$, $m = poly(n)\geq n$.  Suppose further that the first four bits of the output stream $C$ obtained by the adversary are $0, 1, 0, 0$, and that the tent map $f$ has parameter $\mu=2$.  Then, given $G$ and a segment of the output stream $C$, the adversary can efficiently obtain the secret key (initial condition) $x_0$, as we show below.

\begin{remark}
If $m<n$ for an $n$ bit key (secret initial condition), then the adversary does not have enough information to uniquely determine the secret.
\end{remark}

\begin{figure}[t]\label{figIters}
\noindent\null\hfill\includegraphics[scale=0.23]{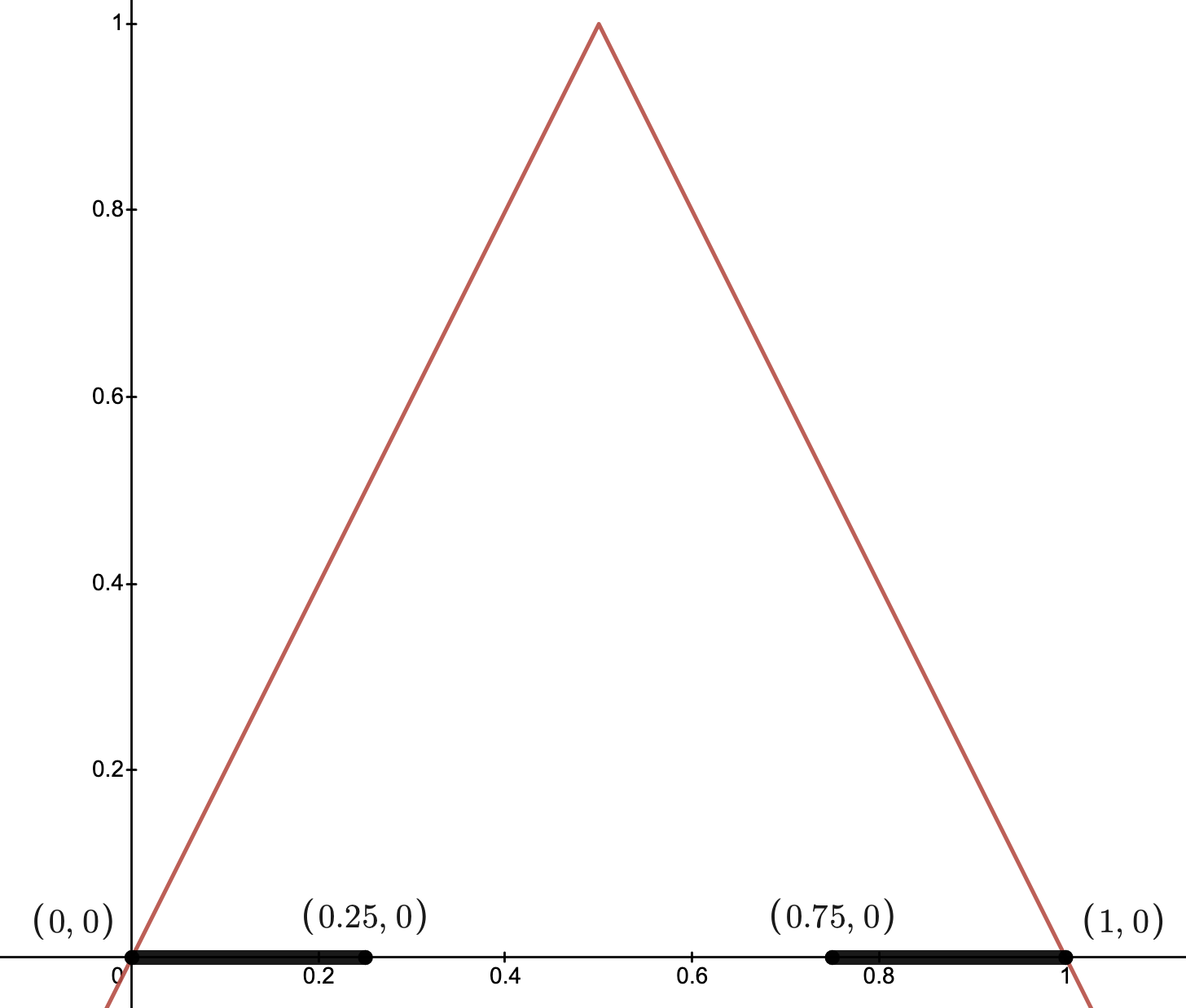} \hfill
\includegraphics[scale=0.23]{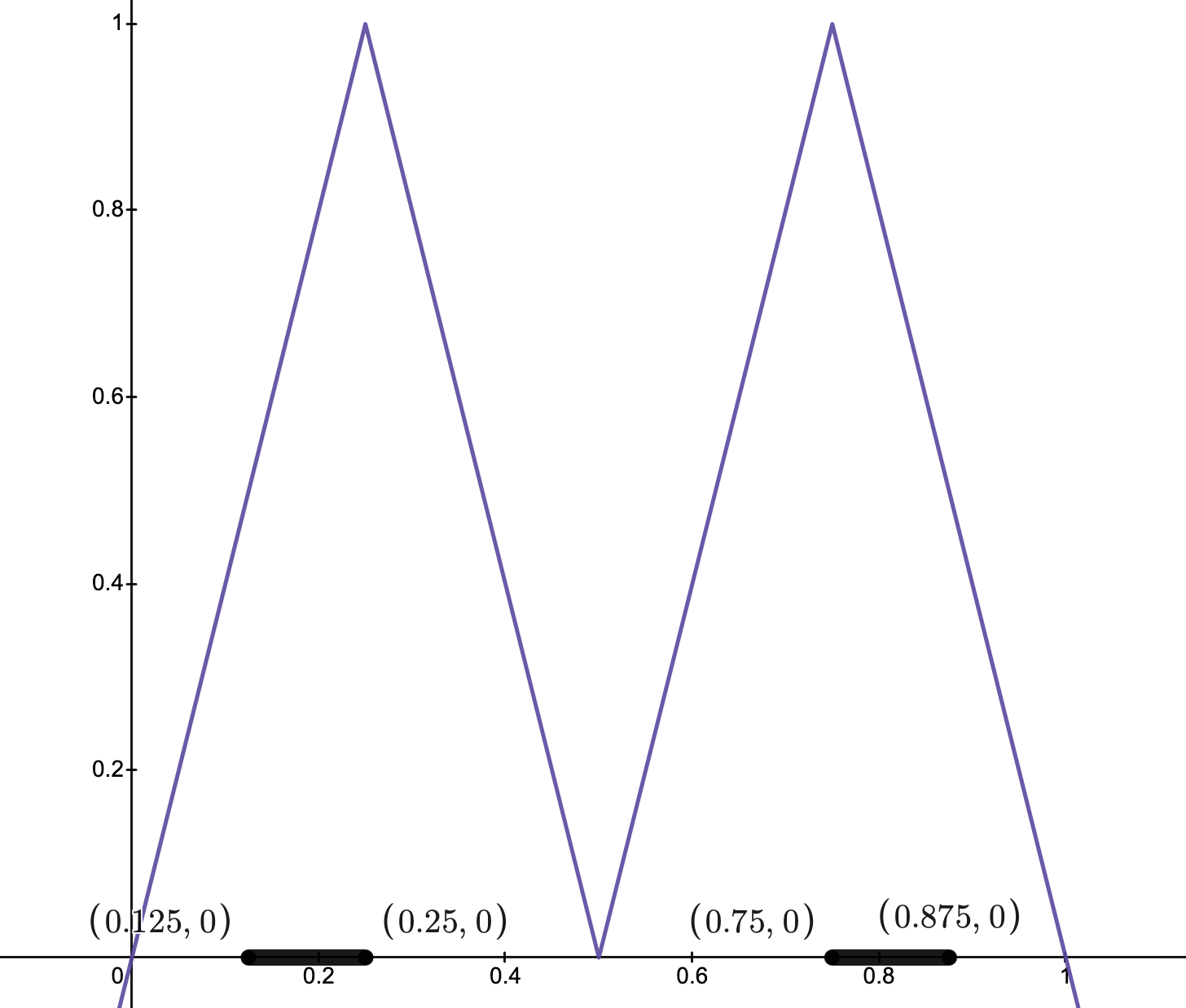} \hfill\null

\noindent\null\hfill\includegraphics[scale=0.23]{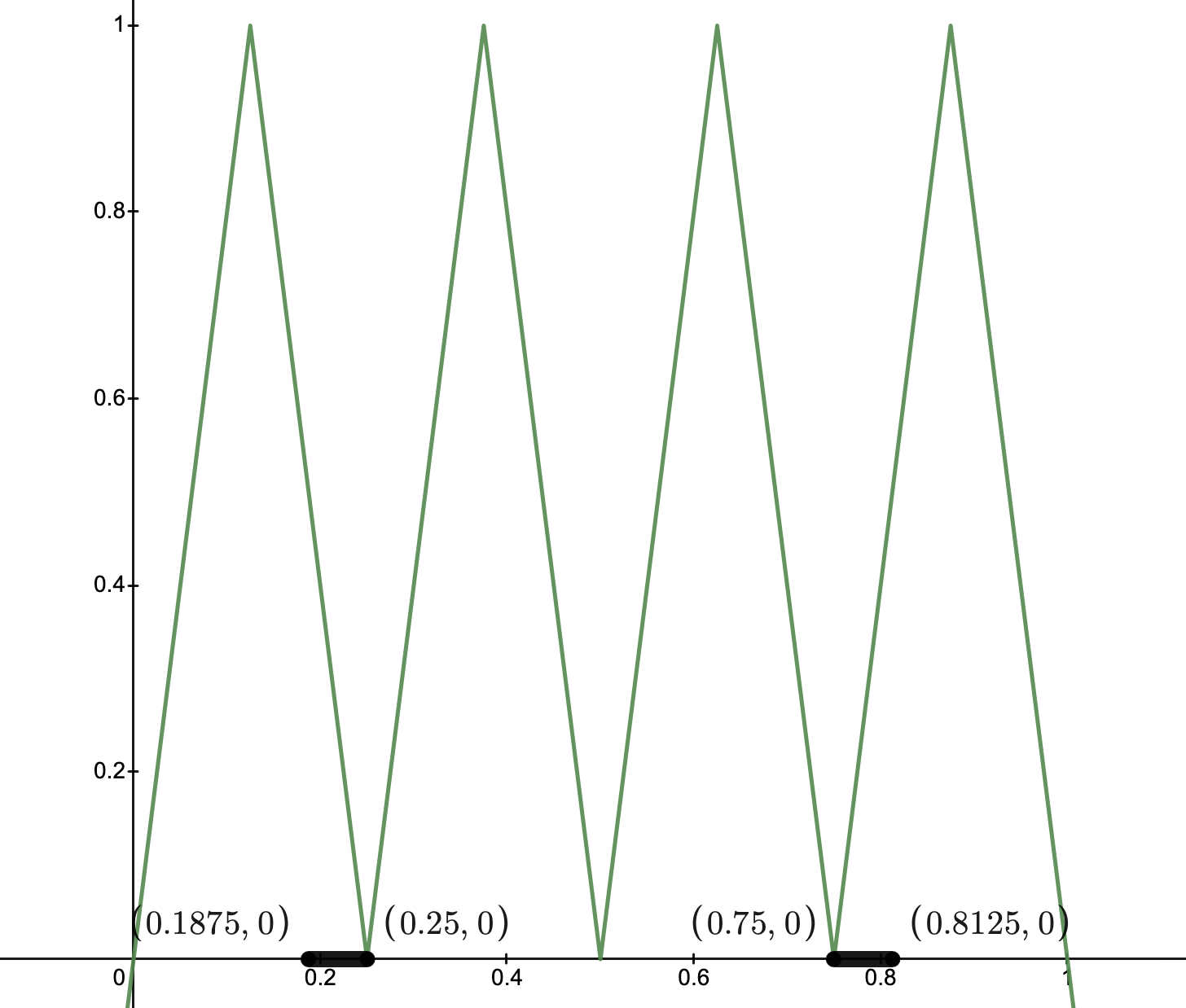} \hfill
\includegraphics[scale=0.23]{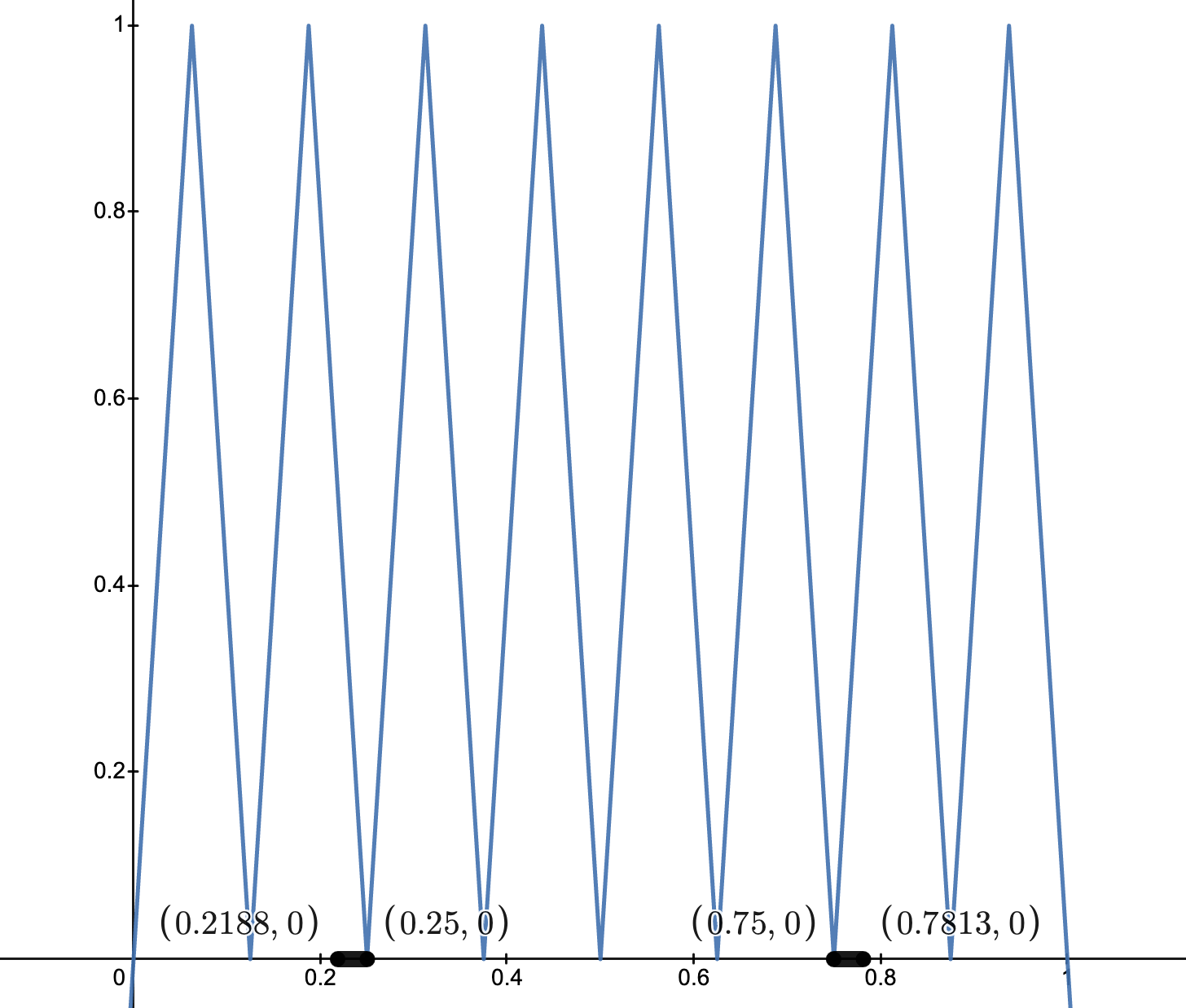} \hfill\null
\caption{The graph on the top left represents the first iteration of the tent map, with the black intervals representing the points $A^{(1)}$ of $[0,1]$ such that $g(f[A^{(1)}]) = c_1 = 0$.  The graph on the top right represents the second iteration of the tent map, $f^2$, with the black intervals representing the subset $A^{(2)}$ of $A^{(1)}$ such that $g(f^2[A^{(2)}]) = c_2 = 1$.  Similarly, the bottom left and right graphs represents the third and fourth iterates of the tent map with the black intervals representing the subsets $A^{(3)}, A^{(4)}$ with $A^{(4)} \subset A^{(3)} \subset A^{(2)} \subset A^{(1)}$ such that $g(f^3[A^{(3)}]) = c_3 =0$ and $g(f^4[A^{(4)}]) = c_4 = 0$.
Note that the length of the black intervals representing candidate secret initial conditions is halved with each iteration of the algorithm.}
\end{figure}

In the steps detailed below, we aim to find a $y \in [0, 1]$ such that $\{G^i(y)\}_{i=1}^m = C$.  Then, given $y$, we (the adversary) can recover all forward iterates, and as we will show in this specific example, all backward iterates as well.  The first four steps of the attack begin as follows.  Recalling that w.l.o.g. we took $c_1=0, c_2=1, c_3=0, c_4=0$.  Refer to Figure \ref{figIters} for visual demonstration of how the proposed attack shrinks the (black) candidate intervals containing $y$ at an exponential rate.
\begin{enumerate}
\item We wish to find the endpoints of interval(s) $A_i^{(1)}$ of $I=[0, 1]$ such that $G(y) = c_1=0$ for all $y \in A_i^{(1)}$.  In our example these sets are $A_1^{(1)}=[0, 0.25]$ and $A_2^{(1)}=[0.75, 1]$ are exactly those intervals such that $G(y)= c_1=0$ for all $y \in A_1^{(1)} \cup A_2^{(1)}$.  Hence, $E_1 = \{(0, 0.25), (0.75, 1)\}$.
\item Compute the endpoints of intervals $A_1^{(2)}\subset A_1^{(1)}$ and $A_2^{(2)}\subset A_2^{(1)}$ such that $G^2[A_1^{(2)}] = G^2[A_2^{(2)}] = c_2 = 1$.  The updated intervals become $A_1^{(2)} = (0.125, 0.25)$ and $A_2^{(2)} = (0.75, 0.875)$, and $E_2 =\{(0.125, 0.25),(0.75, 0.875)\}$.
Again, using the continuity of $f$ computing such endpoints efficiently becomes trivial by leveraging either a binary search or bisection method after subtracting $1/2$ from $f^2(x)$.
\item Compute the endpoints of intervals $A_1^{(3)}\subset A_1^{(2)}$ and $A_2^{(3)}\subset A_2^{(2)}$ such that $G^3[A_1^{(3)}] = G^3[A_2^{(3)}] = c_3 = 0$.  The updated intervals become $A_1^{(3)} = (0.1875, 0.25)$ and $A_2^{(3)} = (0.75, 0.8125)$, and $E_3 = \{(0.1875, 0.25)(0.75, 0.8125)\}$.
\item Compute the endpoints of intervals $A_1^{(4)}\subset A_1^{(3)}$ and $A_2^{(4)}\subset A_2^{(3)}$ such that $G^4[A_1^{(4)}] = G^4[A_2^{(4)}] = c_4 = 0$.  The updated intervals become $A_1^{(4)} = (0.2188, 0.25)$ and $A_2^{(4)} = (0.75, 0.7813)$, and $E_4 = \{(0.2188, 0.25),(0.75, 0.7813)\}$.
\item We now see that it is only the points $y$ in $A_1^{(4)} \cup A_2^{(4)}$ that satisfy $G(y)=0, G^2(y)=1, G^3(y)=0$ and $G^4(y)=0$.  Continuing on in this fashion we can rapidly obtain a small set of candidate initial conditions that allow us to break the stream cipher and decrypt all future messages.
\end{enumerate}

\begin{remark}
Note that in the above example, the length of each interval of candidate points decreases by a factor of $1/2$ with each iteration.  As such, in this case, after just $n+1$ iterations of the algorithm, a unique value $y$ such that $\{G^i(y)\}_{i=1}^m = C$ will be found.
\end{remark}

Note that in the case of \cite{li2017image} as well as many others, given the internal state of the stream cipher $G^j(x_0)$ for some $j>0$, we can determine the initial condition $x_0$.  In the case of our example with the tent-map parameter fixed at $\mu=2$, determining $x_0$ becomes trivial once we have obtained a $y \in [0, 1]$ such that $f^j(x_0) = y$ as outlined in the above attack: $f$ maps $I = \{i/2^l\}_{i=0}^{2^l} \subset [0, 1] $ to itself.  As such, when the parameter of $f$ is $\mu=2$, for any $x \in I$, $f^{-1}(x) = \{p, q\}$ where $p$ and $q$ have form $j/2^l, k/2^l$ where either $j$ is even and $k$ is odd or vice versa. $f$ is a $2-1$ mapping, and takes all inputs in $I$ to a point in $I$ with even numerator, so given $f^j(x_0) = y$, when inverting $f$ simply choose the preimages that have an even numerator.

More generally however, the chaotic maps that underly many chaos-based cryptosystems are $2$ to $1$ maps of the interval, and inverting these maps to find ``true" initial conditions is similarly easy by leveraging numerical errors.  Namely, given $f^j(x_0) = x_j$, compute the pair of points $\{x_{i_1}, x_{i_2}\} = f^{-1}(x_j)$, and check if $f(x_{i_1}) = x_j$ and if $f(x_{i_2}) = x_j$.  If so, recurse and run the same operation for both points.  If not, i.e. if $f(x_{i_1})\neq x_j$, then $x_{i_1}$ cannot be a ``true" preimage of $x_j$, so ignore this point and recurse on $x_{i_2}$.  Continue in this manner for $j$ steps until either a unique point $x_0$ is found, or a small set of candidate initial conditions is found.  

Indeed, numerical experiment indicates that roundoff error quickly reveals what the true preimages were, thereby making schemes like that introduced in \cite{kotulski1} impossible to securely implement without serious modification (see \cite{nardo2021reliable} for an example attempt to remedy this problem and \cite{li2001pseudo} and the references therein for discussion of the problem).

\begin{remark}
The stream cipher described in \cite{li2017image} does not explicitly keep the parameter $\mu$ of the tent map secret.  However, even if it were kept secret, as it is in many related encryption schemes, the algorithm introduced here can still be used to break the schemes.  This is due to the fact that the tent map is only surjective when $\mu=2$ and exhibits ideal chaotic properties for parameter values very close to 2 (see \cite{patidar} for discussion).  As such, there is a small range of parameters $\mu$ that give the necessary behavior, leaving the key-space vulnerable to brute-force attacks.  
Additionally, by leveraging the continuity of the parameter, an algorithm using the basic logic of the algorithm given here can be devised which simultaneously finds both the secret initial condition and parameter, however further discussion is out of the scope of this paper.
\end{remark}

\section{Hard problems in dynamics}\label{sec3}

The attack outlined in the previous section, along with the fact that modern cryptography focuses on the construction of cryptosystems based on a few well understood problems, points to the fact that efforts in chaos-based cryptography should be focused not on creating new schemes but on developing candidate one-way functions (OWFs), and then trying to find ways to efficiently invert such candidate OWFs. If no such algorithm is found, then this would be good evidence that the problem at hand is a OWF.

In this section, we give one such candidate one-way function.  The problem introduced here does not appear to be susceptible to many well-known attacks, including that given in this paper, and is a generalization of the problems introduced in \cite{li2001pseudo} and \cite{patidar}, which, to the best of the author's knowledge, have not been broken.  In both \cite{li2001pseudo} and \cite{patidar}, the authors describe a pseudorandom generator based on the iteration of two chaotic maps running side-by-side and starting from different initial conditions, and the pseudorandom bit sequence is generated by comparing the outputs of both the chaotic maps.  The problem we introduce here is of a similar nature.

\subsection{A possibly hard problem in discrete dynamics.}

Let $f:[0, 1]\rightarrow [0, 1]$ be a surjective, chaotic map of the interval.  Let $x_0, x_1,...,x_{N-1} \in \{0, 1\}^n$ be randomly chosen initial conditions treated as decimals in $[0, 1]$, and let $N = 2^m$ for $m\geq 1$.  Then define the map $h$ as
\begin{equation}
h(x_0,...,x_{N-1}) = \begin{cases}
0 \text{ if } f(x_0) = \min\{f(x_0), f(x_1),...,f(x_{N-1})\},\\
1 \text{ if } f(x_1) = \min\{f(x_0), f(x_1),...,f(x_{N-1})\},\\
\quad \vdots\\
N-1 \text{ if } f(x_{N-1}) = \min\{f(x_0), f(x_1),...,f(x_{N-1})\}.
\end{cases}
\end{equation}
where we take $h(x_0, x_1,...,x_{N-1}) \in \{0, 1\}^m$.  Moreover, we define the $i$th iterate of $h$ to be
\begin{equation}
h^i(x_0,...,x_{N-1}) = \begin{cases}
0 &\text{ if } f^i(x_0) = \min\{f^i(x_0), f^i(x_1),...,f^i(x_{N-1})\},\\
1 &\text{ if } f^i(x_1) = \min\{f^i(x_0), f^i(x_1),...,f^i(x_{N-1})\},\\
\vdots\\
N-1 &\text{ if } f^i(x_{N-1}) = \min\{f^i(x_0), f^i(x_1),...,f^i(x_{N-1})\}
\end{cases}
\end{equation}
for all $i\geq 1$.  A candidate one-way function based on chaotic dynamics is then as follows.

\begin{problem}\label{prob1}
Let $l = poly(n N/m) > n N/m$, and let $\textbf{x} = (x_0, x_1,...,x_{N-1})$ be sampled uniformly at random.  Then, given a length $m l$ binary string $h(\textbf{x})h^2(\textbf{x})\cdots h^l(\textbf{x})$, along with the description of $h$ and $f$, compute $\textbf{x}$.
\end{problem}

Given Problem \ref{prob1}, we have the following open question:

\begin{question*}\label{question1}
Does there exist an algorithm running in time $poly(nN)$ that solves Problem \ref{prob1} with non-negligible probability?
\end{question*}

If sufficient effort is put into deciding whether the answer to this question is ``yes" or ``no," and no-one finds an algorithm to solve Problem \ref{prob1} efficiently, then Problem \ref{prob1} can be taken to be a good candidate for a one-way function, and those designing chaos based cryptosystems can construct provably secure schemes such as stream ciphers or block ciphers under the assumption that Problem \ref{prob1} has a negative answer.

We comment that there is already good evidence that Problem \ref{prob1} is computationally intractable.  This evidence comes from the work of \cite{li2001pseudo} and \cite{patidar}, who study the problem in the $N = 2$ case, and give a variety of statistical tests (including the DIEHARD and NIST statistical test suites) and other evidence such as theorems pointing to the fact that Problem \ref{prob1} is hard, and hence a good candidate one-way function.

Natural next steps toward putting chaos-based cryptography on a rigorous and modern foothold would include making a serious effort to answer Question \ref{question1}, and then developing schemes proven to be secure under the assumption that Question \ref{question1}, or similar problems have a negative answer, as well as determining how such problems can be most efficiently implemented.
It would also be worth considering related problems to that given above.  For example, in the paper \cite{applebaum2010cryptography}, the authors introduce a related ``inversion problem" for cellular automata.  This problem can be naturally extended from cellular automata to general dynamical systems: given a target state, determine the initial condition from a list of possible initial conditions. This problem is also a good candidate for a one-way function based on chaotic dynamics.  Indeed, in \cite{applebaum2010cryptography}, the authors show that the inversion problem for cellular automata is \textsf{NP}-hard to solve in the worst case, which gives extremely strong evidence that it is hard. 

Finally, perhaps one of the most interesting and possibly fruitful directions of study would be finding ``trap-door" functions in discrete dynamical systems that can be used to construct public-key cryptosystems (see \cite{katzlindell} for details).  It is also interesting to consider how a deeper understanding of the computational complexity of dynamical systems problems may lead to a deeper insight into the dynamical systems themselves; much in the same way of how studying the dynamics of a shift map on a symbol space can provide deeper understanding of chaotic maps of the interval for instance.

\bibliography{references}
\end{document}